\documentclass{article}

\usepackage{amssymb,amsmath,xcolor,graphicx,xspace,colortbl,rotating} %
\usepackage[raggedrightboxes]{ragged2e} 
\usepackage{amsmath}  
\usepackage{boxedminipage}  
\usepackage{ragged2e}  
\usepackage{tabulary}  
\usepackage{xcolor}  
\graphicspath{{january18arxiv_graphics/}{january18arxiv_tcache/}{january18arxiv_gcache/}}
\DeclareGraphicsExtensions{.pdf,.eps,.ps,.png,.jpg,.jpeg}

\begin{document}
\title{Completing the physical representation of quantum algorithms provides a quantitative explanation of their computational speedup
}
\author{Giuseppe Castagnoli
 \\
former Elsag Bailey Quantum Laboratory
}
\maketitle
\begin{abstract}The usual representation of quantum algorithms, limited to the process of solving the problem, is physically incomplete. We complete it in three steps: (i) extending the representation to the process of setting the problem, (ii) relativizing the extended representation to the problem solver to whom the problem setting must be concealed, and (iii) symmetrizing the relativized representation for time reversal to represent the reversibility of the underlying physical process. The third steps projects the input state of the relativized representation, where the problem solver is completely ignorant of the setting and thus the solution of the problem, on one where she knows half solution (half of the information specifying it when the solution is an unstructured bit string). Completing the physical representation shows that the number of computation steps (oracle queries) required to solve any oracle problem in an optimal quantum way should be that of a classical algorithm endowed with the advanced knowledge of half solution. This fits the major quantum algorithms known today and would solve the quantum query complexity problem.                                                                                            
\end{abstract}

\section{Introduction
}
          The \textit{quantum computational speedup} is the fact that quantum algorithms solve the respective problems with fewer computation
steps (\textit{oracle queries} in the case of \textit{oracle problems}) than their best classical counterparts, sometimes fewer than classically possible.

A paradigmatic
example is the simplest instance of the quantum algorithm devised by Grover $\left [1\right ]$. Bob, the problem setter, hides a ball in one of four boxes -- drawers from now on.
Alice, the problem solver, is to locate it by opening drawers. In the classical case, she needs to open up to three drawers, always one in the quantum case.

The drawer and ball problem is an example of oracle problem. The operation of checking whether the ball is in a drawer is an example of oracle query, or \textit{function evaluation}. Bob chooses a value of $b$ (hides the ball in drawer $b$) and gives Alice a black box that computes the Kronecker function $\delta \left (b ,a\right )$; Alice checks whether the ball is in drawer $a$ by evaluating $\delta \left (b ,a\right )$. Most quantum algorithms solve oracle problems.

More in general, Bob chooses a function from a set of functions and gives Alice the black box that computes it. Alice, who knows the set of functions but not Bob's choice, is to find a characteristic of the function computed by the black box (the value of $b$ in the drawers and ball problem) by performing function evaluations.

The quantum speedup comes from comparing the number of function evaluations required to solve the oracle problem quantumly and classically. Dozens of different speedups have been discovered.

It should be noted that each quantum algorithm has been found by means of ingenuity. In mainstream literature, there is no fundamental
explanation of the speedup, no unified quantitative explanation of it. This seems to be a lacuna of quantum computer science. Quantum cryptography, the other pillar of quantum information, directly relies on the foundations of quantum mechanics -- e. g. on the EPR situation $\left [2\right ]$.

Here we ascribe our limited understanding of the speedup to the fact that
the physical representation of quantum algorithms, limited to the input-output transformation typical of computation, is physically
incomplete. It consists of a unitary transformation followed by the final measurement required to read the solution. As well known, the complete representation
of a quantum process must include the initial measurement, the unitary transformation of the state after measurement, and the final measurement. 

We show that just completing the physical representation of quantum algorithms provides a fundamental, quantitative, explanation of their speedup. This is done in three steps: 

\begin{enumerate}
\item We extend the usual representation to the process of setting the problem (e. g. of setting the number of the drawer with the ball). Initially the problem setting is completely undetermined. By the initial measurement Bob selects a problem setting at random, then he unitarily transforms it into the desired setting. By performing function evaluations interleaved with other suitable unitary transformations, Alice sends the input state prepared by Bob into the output state encoding the solution of the problem, then she acquires the solution by the final measurement. Since this latter measurement leaves the quantum state unaltered, there is a unitary transformation between the initial and final measurement outcomes.

2. The extended representation works for Bob and any external observer, not for Alice. It would tell her the setting and thus the solution of the problem before she performs any function evaluation. To her, this setting must be hidden inside the black box. This is physically represented by postponing the projection of the quantum state associated with the initial measurement at the end of Alice's problem-solving action. As a consequence, in the representation relativized to Alice, the input state of the quantum algorithm becomes one of complete ignorance of the problem setting chosen by Bob.

3. We represent the physical reversibility of the computation process (there is a unitary transformation between the initial and final measurement outcomes) by symmetrizing the two representations for time reversal. This leaves the representation to Bob and any external observer, where no observer is shielded from the result of any measurement as usual, unaltered. It changes that to Alice, who is shielded from the result of the initial measurement. The former input state to her, of complete ignorance of the problem setting, is projected on one where she knows a part of it that corresponds to half solution (half of the information specifying it in the case it is an unstructured bit string). As a consequence, the number of function evaluations required to solve the oracle problem in an optimal quantum way is that required by a classical algorithm endowed with the advanced knowledge of half solution.\end{enumerate}

Preliminary versions of the present explanation of the speedup were already provided in the evolutionary approach $[3 \div 7]$. The main novelty of the present work is deriving the explanation by progressively completing the physical representation of quantum algorithms. This brought us to a more systematic development of our argument and to the following clarifications.

In $\left [5 \div 7\right ]$ we had just highlighted the mathematical possibility that Alice knows in advance any part of the solution of the problem she will read in the future. Her knowing half of it was an empirical fact ascertained for all the major quantum algorithms. Now we have been able to derive the advanced knowledge of half solution from the same temporal symmetry that stands at the basis of time-symmetric quantum mechanics. As a consequence, the unitary part of the quantum algorithm with respect to Alice turns out to be a quantum superposition of time-symmetrized algorithms in each of which she knows in advance one of the possible halves of the solution of the problem she will read in the future. By the way, this picture is mathematically exact; the former picture, that the quantum algorithm is a sum over classical histories in each of which Alice knows in advance half solution, was still ill defined.

This time we also explicitly address the controversial character of the work in an extended discussion.

\section{Completing the representation
}
Let us consider the four drawers instance of Grover algorithm. Say that the number of the drawer with the ball is $01$. The usual representation of the quantum algorithm, limited to Alice's problem-solving action, consists of the input-output transformation
typical of computations, see the following table. \begin{equation}
\setlength\fboxrule{0cm}\setlength\fboxsep{0cm}\fcolorbox[HTML]{FFFFFF}{FFFFFF}{
\begin{tabular}[c]{lll} &  & $meas. 

\hat{A}$ \\
$\left \vert 00\right \rangle _{A}$ & $ \Rightarrow U \Rightarrow $ & $\left \vert 01\right \rangle _{A}$
\end{tabular}
}
\label{usual}
\end{equation}Alice works on a quantum register $A$, of basis vectors $\left \vert 00\right \rangle _{A} ,\left \vert 01\right \rangle _{A} , . . .$ , meant to contain first the argument of function evaluation then the solution of the problem. The initial state of this register should be like a blank blackboard, say it is  $\left \vert 00\right \rangle _{A}$. Alice, by the unitary transformation $U$, sends it into the output state$\left \vert 01\textit{}\right \rangle _{A}$ that encodes the solution of the problem -- follow the horizontal arrows. Here it suffices to know that $U$ involves just one function evaluation, by the way performed on a quantum superposition of all the possible values of the argument (in \textit{quantum parallel computation}). Throughout this work, we will not need to go inside the unitary part of the quantum algorithm.

 Eventually Alice acquires the solution by measuring the content of register $A$, namely the observable $\hat{A}$ of eigenstates $\left \vert 00\right \rangle _{A} ,\left \vert 01\right \rangle _{A} , . . .$ and eigenvalues respectively $00 ,01 , . . .$ The quantum state $\left \vert 01\right \rangle _{A}$, already an eigenstate of $\hat{A}$, remains unaltered.

The representation of table (\ref{usual}) is physically incomplete in two related ways: (i) it lacks the initial measurement and (ii) the number of the drawer with the ball is not represented physically, i. e. by a quantum state (the number in question is implicit in the mathematics of $U$). Both things are irrelevant to the end of describing the quantum algorithm, not to that of understanding the reason for the speedup.\qquad 

\subsection{
Extending the representation
}

The first step of completing the representation is extending it to the process of setting the problem -- see the following table. 

\begin{equation}\begin{array}{ccc}\;\text{meas.}\;\hat{B} & \, & \;\text{meas.}\;\hat{A} \\
\left (\vert 00 \rangle _{B} +\vert 01 \rangle _{B} +\vert 10 \rangle _{B} +\vert 11 \rangle _{B}\right ) \vert 00 \rangle _{A} & \, & \, \\
\Downarrow  & \, & \, \\
\vert 01 \rangle _{B} \vert 00 \rangle _{A} &  \Rightarrow \mathbf{U} \Rightarrow  & \vert 01 \rangle _{B} \vert 01 \rangle _{A}\end{array} \label{ex}
\end{equation}

We add a possibly imaginary register $B$, of basis vectors $\vert 00 \rangle _{B} ,\vert 01 \rangle _{B} ,\ldots $, that contains $b$, the number of the drawer with the ball. We assume that, initially, this register is in a quantum superposition of all the possible
values of $b$ (top-left corner of the table) -- normalization is disregarded. In view of what will follow, we need to start from scratch, from
an initial state where the value of $b$ is completely undetermined.

It would be more appropriate to start with a maximally mixed state of register $B$ as we did in $\left [7\right ]$. We chose the present superposition to simplify the notation. Everything is the same since, in quantum oracle computing, there is never interference between the basis vectors of register $B$. Function evaluations do not change these vectors ($B$ is the control register here) and the other unitary transformations of Alice's problem solving action are the identity on register $B$ (Alice's action never changes the problem setting). 

By the way, this also implies that the reduced density operator of register $B$ remains unaltered through $\mathbf{U}$ and that $\hat{B}$ can be measured at any time between the beginning and the end of it.

Bob measures the content of register $B$, namely the observable $\hat{B}$ of eigenstates  $\left \vert 00\right \rangle _{B} ,\left \vert 01\right \rangle _{B} , . . .$ and eigenvalues respectively $00 ,01 , . . .$ (note that $\hat{B}$ and $\hat{A}$ commute). This projects the quantum superposition on an eigenstate of $\hat{B}$ selected at random, say $\left \vert 01\right \rangle _{B}$ -- follow the vertical arrow. Ordinarily, Bob would unitarily change the number selected at random into the desired number. We jump this transformation for simplicity. To the end of assessing the number of function evaluations required to solve the problem quantumly, the fact that the problem setting is selected at random or intentionally is irrelevant.

Up to this point Bob has prepared the input state of the quantum algorithm. Alice unitarily sends it into the output state $\left \vert 01\right \rangle _{B}\left \vert 01\right \rangle _{A}$, where register $A$ contains the solution -- follow the horizontal arrows. Note that the unitary transformation $\mathbf{U}$ is different from $U$; it does not contain any information about the number of the drawer with the ball selected by Bob -- this information is now in register $B$. Eventually Alice measures $\hat{A}$, thus acquiring the solution of the problem. This leaves the quantum state unaltered -- the output state of register $A$ is already an eigenstate of $\hat{A}$. There is thus a unitary transformation between the initial and final measurement outcomes, the process between them is physically reversible (no information is destroyed along it).

\subsection{
Relativizing the extended representation to Alice
}
The extension of the quantum algorithm to the process of setting the problem faces us with a novel situation. The outcome of the initial measurement, $\left \vert 01\right \rangle _{B}\left \vert 00\right \rangle _{A}$, of course tells Bob that the number of the drawer with the ball is $01$. The point is that, if $\left \vert 01\right \rangle _{B}\left \vert 00\right \rangle _{A}$ was also the state to Alice, it would tell her the number of the drawer with the ball before she begins her problem solving action. 

To Alice, the number $01$ should be hidden inside the black box that performs the function evaluations. Since this notion is essential from the computational standpoint (see Section 4), we argue that, in the quantum case, it should be represented physically. As noted by Deutsch in his seminal 1985 paper $\left [8\right ]$, the all with quantum computation is representing abstract computational notions physically.

 To conceal to Alice the number of the drawer with the ball selected by Bob, it suffices to postpone  at the end of the unitary part of her problem-solving action the projection of the quantum state associated with the initial measurement of $\hat{B}$ (correspondingly, the two end states of the projection are propagated by $\mathbf{U}$). We note that we are just exploiting a degree of freedom of the quantum description: as well known, the projection associated with a quantum measurement can be postponed at will along a unitary transformation that follows it. 

Moreover, since  $\hat{B}$ and $\hat{A}$ are commuting observables, we can postpone the projection associated with the initial measurement after that associated with the final one. In fact, all is as if the very measurement of $\hat{B}$ was postponed, and we are free of postponing it immediately before or after that of $\hat{A}$. With this, Alice's view of the extended representation becomes:

\begin{equation}
\setlength\fboxrule{0cm}\setlength\fboxsep{0cm}\fcolorbox[HTML]{FFFFFF}{FFFFFF}{
\begin{tabular}[c]{lll}meas.$

\hat{B}$ &  & meas. of $\hat{A}$ \\
$\left (\left \vert 00\right \rangle _{B} +\left \vert 01\right \rangle _{B} +\left \vert 10\right \rangle _{B} +\left \vert 11\right \rangle _{B}\right )\left \vert 00\right \rangle _{A}$ & $ \Rightarrow \mathbf{U} \Rightarrow $ & $\left \vert 00\right \rangle _{B}\left \vert 00\right \rangle _{A} +\left \vert 01\right \rangle _{B}\left \vert 01\right \rangle _{A} +\left \vert 10\right \rangle _{B}\left \vert 10\right \rangle _{A} +\left \vert 11\right \rangle _{B}\left \vert 11\right \rangle _{A}$ \\
 &  & $\Downarrow $ \\
 &  & $\left \vert 01\right \rangle _{B}\left \vert 01\right \rangle _{A}$
\end{tabular}
}
\label{alice}
\end{equation}

We can see that the initial measurement of $\hat{B}$ leaves the initial state, of complete indetermination of the problem setting, unaltered; in the input state of the quantum algorithm, Alice remains completely ignorant of the number of the drawer with the ball selected by Bob. Under $\mathbf{U}$ (horizontal arrows), this input state evolves into the quantum superposition of four tensor products, each the product of a number of the drawer with the ball and the corresponding solution (that same number but in register $A$). Then Alice measures $\hat{A}$ projecting the superposition on the number of the drawer with the ball already selected by Bob (vertical arrow). This particular projection is unpredictable by Alice as usual, it is already known by Bob (and any external observer). Note that it also coincides with the projection due to the initial measurement of $\hat{B}$ postponed at the end of Alice's action.

We call this second representation of the extended quantum algorithm the representation relativized to Alice. It should be noted that, implicitly, we have resorted to relational quantum mechanics $\left [9 \div 11\right ]$ according to which quantum states are observer dependent -- see Section 4.1.

\subsection{Symmetrizing for time reversal
}
 Let us recall the fact that there is the unitary transformation $\mathbf{U}$ between the initial and final measurement outcomes. Mathematically, the selection of the random outcome of the initial measurement (the number of the drawer with the ball $01$) can be ascribed indifferently to the initial measurement of $\hat{B}$ or the final measurement of $\hat{A}$. In the former case, the initial measurement outcome should be propagated forward in time by $\mathbf{U}$ as usual, until it becomes the outcome of the final measurement. In the latter, we should propagate the final measurement outcome backward in time by the inverse of $\mathbf{U}$, until it becomes the outcome of the initial measurement. Then the question becomes: which measurement performs the selection? 

Our answer is: none of them alone, since this would introduce a preferred direction of time incompatible with the reversibility of the quantum process. Moreover, we should require that the selection of the random outcome of the initial measurement evenly shares between the initial and final measurements. Otherwise there would still be a time asymmetry.

We should ascribe the selection of half of the information that specifies the random outcome of the initial measurement to the initial measurement, that of the complementary half to the final measurement. By the way, we are applying Occam razor, which here becomes the principle that there should be no redundancies between the selections performed by the two measurements (the same information should not be selected twice). As there are many ways of taking half of the information, all these ways should be taken in quantum superposition; how, will be more easily explained further below.

In order that the information about the random outcome of the initial measurement is shared evenly, the initial and final measurements, in presence of one another (contextually), should reduce to partial measurements that evenly and without redundancies contribute to the selection of the number of the drawer with the ball.

An example is as follows. With random outcome of the initial measurement $01$, we can assume that things went as follows: (i) the initial measurement, reduced to that of $\hat{B}_{l}$ (the left digit of the number contained in register $B$), selected the $0$ of $01$ and (ii) the final measurement, reduced to that of $\hat{A_{r}}$ (the right digit of the number contained in register $A$), selected the $1$ of $01$. The selection performed by the initial measurement should be propagated forward in time by $\mathbf{U}$, that performed by the final measurement backward in time by the inverse of $\mathbf{U}$. Performing these two propagations in a sequence time-symmetrizes the quantum process. We will see that this is without consequences in the representation with respect to Bob and any external observer, where no observer is shielded from the outcome of any measurement as usual. It explains the speedup in the representation with respect to Alice, who is shielded from the outcome of the initial measurement.

\subsubsection{
Symmetrizing for time reversal the representation to Bob and any external observer
}
In the case of the extended representation (that to Bob and any external observer), the symmetrization
for time reversal is described by the following table.

\begin{equation}\begin{array}{ccc}\;\text{meas. of}\;\hat{B}_{l} & \, & \;\text{meas. of}\;\hat{A_{r}} \\
\left (\vert 00 \rangle _{B} +\vert 01 \rangle _{B} +\vert 10 \rangle _{B} +\vert 11 \rangle _{B}\right ) \vert 00 \rangle _{A} & \, & \, \\
\Downarrow  & \, & \, \\
\, & \, & \, \\
\left (\vert 00 \rangle _{B} +\vert 01 \rangle _{B}\right ) \vert 00 \rangle _{A} &  \Rightarrow \mathbf{U} \Rightarrow  & \vert 00 \rangle _{B} \vert 00 \rangle _{A} +\vert 01 \rangle _{B} \vert 01 \rangle _{A} \\
\, & \, & \Downarrow  \\
\vert 01 \rangle _{B} \vert 00 \rangle _{A} &  \Leftarrow \mathbf{U}^{\dag } \Leftarrow  & \vert 01 \rangle _{B} \vert 01 \rangle _{A}\end{array} \label{syb}
\end{equation}

The initial measurement of $\hat{B}_{l}$, selecting the $0$ of $01$, projects the initial quantum superposition (vertical arrow on the top left of the table) on the superposition of the terms
beginning with $0$, yielding a new input state of the quantum algorithm. Under $\mathbf{U}$ (horizontal arrows), this state evolves into the superposition of the two corresponding products \textit{number of the drawer with the ball}$ \times $\textit{corresponding solution}. Then the final measurement of $\hat{A_{r}}$, selecting the $1$ of $01$, projects the superposition in question (vertical arrow) on the term ending in $1$, namely on the original output state of the quantum algorithm to Bob and any external observer. Propagating backward in time
this output state, by $\mathbf{U}^{\dag }$ (left looking horizontal arrows), rebuilds the original input state.

The bottom line of table (\ref{syb}), where of course we can replace $ \Leftarrow \mathbf{U}^{\dag } \Leftarrow $ by $ \Rightarrow \mathbf{U} \Rightarrow $, is the symmetrized representation. We can see that the quantum process between  the initial and final measurement outcomes is the same of the original representation.

The representation to Bob and any external observer, which is the ordinary representation of a reversible quantum process where no observer is shielded from the outcome of any measurement, is natively invariant with respect to the present kind of time-symmetrization.

\subsubsection{Symmetrizing for time reversal the representation
relativized to Alice
}
We will see that symmetrization for time reversal highlights a particular internal structure of the representation relativized
to Alice.
\begin{equation}\begin{array}{ccc}\;\text{meas. of}\;\hat{B}_{l} & \, & \;\text{meas. of}\;\hat{A_{r}} \\
\left (\vert 00 \rangle _{B} +\vert 01 \rangle _{B} +\vert 10 \rangle _{B} +\vert 11 \rangle _{B}\right ) \vert 00 \rangle _{A} &  \Rightarrow \mathbf{U} \Rightarrow  & \vert 00 \rangle _{B} \vert 00 \rangle _{A} +\vert 01 \rangle _{B} \vert 01 \rangle _{A} +\vert 10 \rangle _{B} \vert 10 \rangle _{A} +\vert 11 \rangle _{B} \vert 11 \rangle _{A} \\
\, & \, & \Downarrow  \\
\left (\vert 01 \rangle _{B} +\vert 11 \rangle _{B}\right ) \vert 00 \rangle _{A} &  \Leftarrow \mathbf{U}^{\dag } \Leftarrow  & \vert 01 \rangle _{B} \vert 01 \rangle _{A} +\vert 11 \rangle _{B} \vert 11 \rangle _{A}\end{array} \label{sya}
\end{equation}In table (\ref{sya}), the projection of the quantum state associated with the initial measurement of   $\hat{B}_{l}$ must be postponed at the end of Alice's problem solving action -- to Alice any information about the problem setting is hidden inside the black box. The input and output states of the forward propagation to Alice (top line of table \ref{sya}) are thus the same of table (\ref{alice}). The measurement of $\hat{A_{r}}$ in the output state, selecting the $1$ of $01$, projects it (vertical arrow) on the superposition of the terms ending in $1$. Propagating this superposition backward in time by $\mathbf{U}^{\dag }$  (left looking horizontal arrows), yields an instance of the time-symmetrized representation to Alice -- bottom line of table (\ref{sya}). We will see in the next section that the superposition of all these symmetrization instances gives back the unitary part of the quantum algorithm to Alice.

\subsection{Interpretation
}
The bottom line of  table (\ref{sya}) is the time-symmetrized representation of the quantum algorithm to Alice in the particular case that the outcome of the initial measurement is $01$ and that the initial and final measurements reduce to those of $\hat{B}_{l}$ and $\hat{A_{r}}$. We repeat it here for convenience with $ \Leftarrow \mathbf{U}^{\dag } \Leftarrow $ replaced by $ \Rightarrow \mathbf{U} \Rightarrow $:  \begin{equation}(\left \vert 01\right \rangle _{B} +\left \vert 11\right \rangle _{B})\left \vert 00\right \rangle _{A} \Rightarrow \mathbf{U} \Rightarrow \left \vert 01\right \rangle _{B}\left \vert 01\right \rangle _{A} +\left \vert 11\right \rangle _{B}\left \vert 11\right \rangle _{A} . \label{distance}
\end{equation}

In total there are three ways of evenly sharing the selection of $01$ between the two measurements, as many as the number of ways of pairing $01$ with another drawer number; the other two are thus:

\begin{equation*}(\left \vert 00\right \rangle _{B} +\left \vert 01\right \rangle _{B})\left \vert 00\right \rangle _{A} \Rightarrow \mathbf{U} \Rightarrow \left \vert 00\right \rangle _{B}\left \vert 00\right \rangle _{A} +\left \vert 01\right \rangle _{B}\left \vert 01\right \rangle _{A}
\end{equation*}\begin{equation*}(\left \vert 01\right \rangle _{B} +\left \vert 10\right \rangle _{B})\left \vert 00\right \rangle _{A} \Rightarrow \mathbf{U} \Rightarrow \left \vert 01\right \rangle _{B}\left \vert 01\right \rangle _{A} +\left \vert 10\right \rangle _{B}\left \vert 10\right \rangle _{A} .
\end{equation*}

On can see that taking the quantum superposing of all the possible time-symmetrization instances for all the possible problem settings rebuilds the unitary part of the quantum algorithm to Alice.

Each time-symmetrization instance of the quantum superposition to Alice changes the former input state to her, of maximal ignorance of the number of the drawer with the ball, into one where she knows that it is is in one of two drawers, either $01$ or $11$ in instance (\ref{distance}). It tells her one of the possible halves of the solution of the problem. 

Correspondingly, what is required of $\mathbf{U}$ is to locate a ball hidden in the pair of drawers $\left \{01 ,11\right \}$ -- this can be seen by comparing the input and output states of equation (\ref{distance}). We call this problem, with reduced computational complexity with respect to the original oracle problem, the \textit{reduced problem}. 

Of course $\mathbf{U}$ is also the unitary transformation that solves the original oracle problem. We are making the basic assumption that the number of function evaluations required by $\mathbf{U}$ should be ascertained on any of the time-symmetrization instances, in fact after completing the physical representation of the quantum algorithm to Alice.  

As any other problem, the reduced problem can always be solved quantumly with the number of function evaluations required to solve it classically -- just one in the present case. This means that $\mathbf{U}$  can be performed with just one function evaluation. To sum up, quantumly, the original problem can always be solved with the number of function evaluations required to solve the reduced problem classically. What we have found until now is an upper bound the the quantum computational complexity of the oracle problem.

In the four drawer case, this upper bound is also a lower one since the problem cannot be solved with less than one function evaluation. To examine the situation in general, let us consider the generic number of drawers $N =2^{n}$. Evenly sharing between the initial and final measurements the selection of the $n$ bits that specify the number of the drawer with the ball selected by Bob implies that the reduced problem is locating the ball in $2^{n/2}$ drawers. This requires classically $\ensuremath{\operatorname*{O}}\left (2^{n/2}\right )$ function evaluations. For what we have seen before, this is also an upper bound to the number of function evaluations required by $\mathbf{U}$. Thus, the original problem of locating the ball in $2^{n}$ drawers can always be solved quantumly with $\ensuremath{\operatorname*{O}}\left (2^{n/2}\right )$ function evaluations.

Now we can ask ourselves whether it could be solved with less function evaluations. Consistently with present assumptions, the answer must be negative. Since Alice is shielded from the outcome of the initial measurement, a further reduction of the number of function evaluations required to solve the oracle problem would imply that more than $n/2$ bits of information about the solution come back in time to her from the final measurement. The process between the initial and final measurement outcomes would no more be time-symmetric.

In the end, the number of function evaluations required to solve the original problem quantumly must be the number required to solve the corresponding reduced problem classically, namely  $\ensuremath{\operatorname*{O}}\left (2^{n/2}\right )$. Note that this must be the number required in the optimal case, a non optimal quantum algorithm could take any higher number of function evaluations. As shown by Long $\left [12 ,13\right ]$, Grover's problem can be solved quantumly with any number of function evaluations equal or above that required by the optimal Grover algorithm.

Let us summarize things. Bob measures $\hat{B}$, projecting the initial state of register $B$ on, say, $\left \vert 01\right \rangle _{B}$. The representation of the quantum algorithm to Alice is thus that of table (\ref{alice}), where the projection on $\left \vert 01\right \rangle _{B}$ is postponed at the end of her problem solving action. Up to this point, the representation to Alice ignores the selection performed by Bob. Its unitary part can be seen as a superposition of time-symmetrized quantum algorithms (instances). In each instance, there is information going back in time from the outcome of the final measurement to that of the initial measurement; Alice knows in advance, before beginning her problem solving action, a part of the problem setting corresponding to half solution. The computational complexity of the problem to be solved by her is correspondingly reduced. Our assumption is that any such instance tells us the number of function evaluations required to solve the problem; it is the number required to solve the reduced problem classically. Eventually, the final Alice's measurement selects the problem setting (already selected by Bob) and the corresponding solution.

In view of what will follow, we pinpoint a different way of obtaining Alice's advanced knowledge. Let us remain in the case that the outcome of the initial measurement is $01$ and that the initial and final measurements reduce respectively to those of $\hat{B}_{l}$ and $\hat{A_{r}}$. We note that measuring $\hat{A_{r}}$ in the output state is the same as measuring  $\hat{B}_{r}$ (the right digit of the number contained in register $B$) and that the latter measurement can be performed indifferently in the input state; the result -- selecting the $1$ of $01$ -- is the same. We can thus split the initial measurement of $\hat{B}$ into two partial measurements such that they evenly and without redundancies contribute to the determination of the problem setting and, through $\mathbf{U}$, of the solution. Either partial measurement projects the initial quantum superposition of all the possible problem settings on a smaller superposition, an instance of Alice's advanced knowledge.

It is useful to keep also in mind how to derive the reduced problem from Alice's advanced knowledge. Let $\sigma  \equiv \left \{00 ,01 ,10 ,11\right \}$ be the set of all the possible problem settings and $\sigma _{1 ,3} \equiv \left \{01 ,11\right \}$ the subset of $\sigma $ Alice knows in advance $b$ belongs to. The reduced problem is obtained from the original oracle problem by replacing $\sigma $ by $\sigma _{1 ,3}$.

\section{
Generalization
}
Until now we have worked on the four drawer instance of Grover algorithm. In this section we generalize things. 

\subsection{
Grover algorithm with any number of drawers
}
 As anticipated in Section 2.4, the number of function evaluations required to solve Grover's problem in an optimal quantum way in the case of $2^{n}$ drawers is  $\ensuremath{\operatorname*{O}}\left (2^{n/2}\right )$. This is in agreement with the \textit{quadratic} speedup delivered by Grover algorithm, which is optimal $\left [12 ,13 ,14\right ]$. 

We should note that Grover algorithm, as revisited by Long $\left [13 ,14\right ]$, gives the solution with certainty for any value of $n$. There is thus always the unitary transformation between the initial and final measurement outcomes required by the present explanation of the speedup.

\subsection{
Generic oracle problem
}
In the case of a generic oracle problem, to start with we need to know the input and output states of registers $B$ and $A$ in the representation of the generic quantum algorithm to Alice. Disregarding normalization, they have always the form:

\begin{equation}\left \vert \psi \right \rangle _{IN} =\sum _{b \in \sigma }\left \vert b\right \rangle _{B}\left \vert 0...\right \rangle _{A} , \label{setting}
\end{equation}\begin{equation}\left \vert \psi \right \rangle _{OUT} =\sum _{b \in \sigma }\left \vert b\right \rangle _{B}\left \vert s\left (b\right )\right \rangle _{A} . \label{solution}
\end{equation}$b$ ranges over $\sigma $, the set of all the problem settings, the notation $0...$ stands for a sequence of all zeros (a blank blackboard) the size of the solution, and $s\left (b\right )\text{}$ (a function of $b$) is the solution of the problem\protect\footnote{
Note that Alice knows the function $s\left (b\right )$, the only thing she does not know is the problem setting selected by Bob. 
}. Of course, once selected the value of $b$, the probability of finding the solution $s\left (b\right )$ by measuring $\hat{A}$  in state (\ref{solution}) is one, as required by the present model to have a unitary transformation between the initial and final measurement outcomes.

 Note that, to write states (\ref{setting}) and (\ref{solution}), it is sufficient to know the problem, namely the set of the problem settings and the corresponding solutions. We do not need to know the input-output transformation $\mathbf{U}$ (such that $\mathbf{U}\left \vert \psi \right \rangle _{IN} =\left \vert \psi \right \rangle _{OUT}$), namely the quantum algorithm that solves the problem. Incidentally, in principle there can always be such a unitary transformation, because here the output has a full memory of the input.

We discuss the case that the solution is a many to one function of the problem setting. 

We note that the projection due to the initial measurement of $\hat{B}$, postponed at the end of Alice's problem solving action, eventually tells her the problem setting (the value of $b$) selected by Bob. Therefore, from the computational standpoint, all is as if Alice measured $\hat{B}$ and $\hat{A}$ in the output state; of course the latter measurement is redundant with the former one since the solution is a function of the problem setting. In the end, we should evenly and non redundantly share the selection of the problem setting and the consequent selection of the solution between the initial and final measurement of  $\hat{B}$. 

Since we are free to bring the final measurement of  $\hat{B}$ into the initial state, one can see that everything boils down to splitting the initial measurement of $\hat{B}$  into two partial measurements that evenly and without redundancy contribute to the determination of the solution. Note that this also implies that the contributions to the determination to the problem setting are even (while the converse may not be true when the solution is a many to one function of the problem setting).  Either partial measurement projects the superposition of all the values of $b$ onto a smaller superposition, an instance of Alice's advanced knowledge.

Note that also the contributions to the determination of the solution can be computed without knowing $\mathbf{U}$. In fact the projection of the state of register $B$ due to either partial measurement of the content of this register in the input state consists of a reduction of the range of the summation over $b$, from $\sigma $ to a $\sigma _{i}$, where $i$ labels the partial measurement at stake. This reduction propagates unaltered to the output state, where it determines the contribution of the partial measurement in question to the determination of the solution. This is because the unitary part of Alice's action does not change the reduced density operator of register $B$ and thus any projection thereof. Therefore everything can be computed on the basis of equations (\ref{setting}) and (\ref{solution}), with no need of knowing $\mathbf{U}$.

In $\left [7\right ]$, we have called the present method of calculating Alice's advanced knowledge and thus the number of function evaluations required to solve the problem in an optimal quantum way the \textit{advanced knowledge rule}. 

\subsection{
Application to other quantum algorithms   
}
In              $\left [7\right ]$, we have checked that the advanced knowledge rule explains the speedup of the major quantum algorithms: besides Grover algorithm, the seminal Deutsch algorithm $\left [8\right ]$, Deutsch\&Jozsa algorithm $\left [15\right ]$, Simon algorithm $\left [16\right ]$ and the algorithms of the Abelian hidden subgroup $\left [17\right ]$ -- about a dozen quantum algorithms among which Shor factorization algorithm $\left [18\right ]$. In all these cases, the number of function evaluations foreseen by the advanced knowledge rule coincides with that of the actual quantum algorithm, which is always optimal in character.

As an example, we summarize the application of the subject rule to the problem solved by Deutsch\&Jozsa algorithm. 

There is a set of binary functions $f_{b}\left (a\right )$ that are either constant or balanced. In constant functions, the values of the function are of course either all zeros or all ones; in balanced functions, half of the values are zeros, the other half are ones. The table below provides four of the eight functions for a two bit argument.

\begin{equation}\begin{array}{cccccc}a & f_{0000}\left (a\right ) & f_{1111}\left (a\right ) & f_{0011}\left (a\right ) & f_{1100}\left (a\right ) &  . . . \\
00 & 0 & 1 & 0 & 1 &  . . . \\
01 & 0 & 1 & 0 & 1 &  . . . \\
10 & 0 & 1 & 1 & 0 &  . . . \\
11 & 0 & 1 & 1 & 0 &  . . .\end{array}
\end{equation}

Bob selects one of these functions, namely a value of the problem setting $b$, and gives Alice the black box that computes it. Alice is to find whether the function selected by Bob is constant or balanced by performing function evaluations for suitable values of the argument $a$.

In the classical case, and in the worst case, the number of function evaluations required to solve the problem grows exponentially with $n$, where $n$ is the number of the bits in $a$. In the quantum case just one is required.

We apply now the advanced knowledge rule to determine the number of function evaluations required to solve the problem in an optimal quantum way. First, note that we have chosen as the problem setting $b$ the table of the function, namely the sequence of function values for increasing values of the argument. Let us assume that the problem setting is $b =0011$ (i. e. $f_{0011}\left (00\right ) =0 ,f_{0011}\left (01\right ) =0$, etc.). We can see that there is only one pair of partial measurements of the content of register $B$ satisfying the advanced knowledge rule. One is the measurement of the two left digits of $b$, which are both zero; Alice's advanced knowledge is correspondingly $b \in \left \{0000 ,0011\right \}\text{.}$ The other is that of the two right digits, which are both one; here Alice's advanced knowledge is  $b \in \left \{0011 ,1111\right \}$. In fact, if one partial measurement gave both zeros and ones, it would already tell Alice the solution, namely that the function is balanced. Then the cases are two: If also the other partial measurement told the solution, there would be redundancy between the two partial measurements. If it did not, the two partial measurements would not contribute evenly to the determination of the solution. In either case there would be a violation of the advanced knowledge rule.

One can see that, more in general,  the advanced knowledge rule requires that Alice knows in advance a \textit{good half table}, namely a half table in which all the values of the function are the same. As a consequence, she can always find whether the function is constant or balanced  by performing just one function evaluation for any value of the argument outside the half table.

Summing up, the advanced knowledge rule says that Deutsch\&Jozsa problem can be solved in an optimal quantum way with just one function evaluation. It is the number required by Deutsch\&Jozsa algorithm, which is of course optimal in character.

The application of the advanced knowledge rule to all the other major exponential speedups is very similar. In all cases, knowing in advance one of two parts of the table of the function that evenly and non redundantly contribute to the determination of the solution allows to find the solution with just one function evaluation (against an exponential number thereof in the classical case). All the problems solved by the algorithms in question have this hidden feature.

Let us consider for example the problem of finding the period of a periodic function. It is the problem solved by Simon algorithm and the quantum part of Shor algorithm. Here the two parts of the table should be two halves of the table of a period of the function such that each half does not contain a same value of the function twice. Otherwise this would already tell the period and there would be the same violation of the advanced knowledge rule that we have seen with Deutsch\&Jozsa problem. Then, just one function evaluation, for any value of the argument outside such a half table, is sufficient to identify the period of the function selected by Bob.

It is reasonable to think that the exponential speedups always come out with structured problems where the advanced knowledge of half solution yields an extraordinary advantage. When the problem is not structured, like in the case of quantum search, the speed up would be quadratic.\qquad

\section{
Discussion and conclusion}
Being based on novel physical considerations, the present explanation of the speedup lends itself to controversy. In this section we defend it by addressing possible objections.

A general reason for objection could be the excess structure and complexity of our approach with respect to the usual representation of quantum algorithms. Our answer would be twofold. First, the physical process is naturally more complex if one does not disregard the concealment of the problem setting to the problem solver. Second, we are in uncharted waters and there are many notions that need to be introduced from scratch. Once these notions were taken for granted, the present explanation of the speedup would become much simpler, as we will see in the conclusion.

In the following, first we summarize and discuss the key steps leading to the subject explanation. This should also provide a more unified vision of what has been done. Then we try to position our approach with respect to literature. Eventually we provide our conclusion.

\subsection{
Key steps
}
The usual representation of quantum algorithms is limited to Alice's problem solving action. It thus consists of the unitary input-output transformation required to solve the problem and the final measurement required to read the solution. The initial measurement, unnecessary to describe the action of solving the problem, is missing. Moreover, the problem setting (the value of $b$) is represented only in a mathematical way (it is implicit in $U$), not physically by a quantum state. The leitmotiv of the present work is completing the physical representation.

The first step is extending the usual representation to the process of setting the problem. Under the present foundational standpoint, it is mandatory for two reasons: (i) the complete representation of a quantum process must consist of initial measurement, unitary evolution, and final measurement and (ii) we must represent physically the problem setting selected by Bob. Of course there is no criticality in this step, which is propaedeutic to the others.

We argue that the second step, relativizing the extended representation to Alice, is the necessary consequence of the first. To see this, in the first place we should realize that oracle problems are games between two players\protect\footnote{
We are making reference to game theory, whose object is studying the mathematical models of conflict and cooperation between intelligent rational decision makers.}. Bob selects a function $f_{b}$ from a set of functions known to both players and gives Alice the black box that computes it. Alice, who does not know the function selected by Bob, is to find a characteristic of it by performing function evaluations. What is interesting is the minimum number of function evaluations required to solve the problem. If Alice knew the function selected by Bob, of course she could solve the problem without performing any function evaluation. Therefore the terms of the game imply that the two players have two different perspectives of the situation.

Moving to quantum computation, namely to the physical representation of the abstract computational notions, we have to physically represent the fact that the function selected by Bob is concealed to Alice.

First, it is natural to map the notion of player on that of observer. With the extension of the quantum process, we have necessarily two different observers of it: Bob, who reads the outcome of the initial measurement (the problem setting) and Alice, who reads the outcome of the final measurement (the solution) and cannot know that of the initial measurement until the end of her problem solving action.

Second, the fact that the two observers must have different perspectives of the physical situation, necessarily brings into play the relational quantum mechanics of Rovelli $\left [9 \div 11\right ]$. According to it, a quantum state has meaning to an observer, like in the Copenhagen interpretation. What the relational representation rejects is the notion of absolute, or observer independent, quantum state. 

For example, at a certain time, to \textit{}one observer the state of a quantum system may be collapsed on an eigenstate of the measured observable, while to another observer the system may still appear to be in a superposition of all the eigenstates. This is exactly the present case: the projection of the quantum state associated with the initial measurement of the content of register $B$ must be immediately visible to Bob as usual, to Alice it must be postponed at the end of her problem solving action. 

The original representation of the quantum process thus splits into two relational representations. The original representation becomes that with respect to Bob and any external observer. The second representation is with respect to Alice; here the projection of the quantum state due to the initial measurement of the content of register $B$ is postponed at the end of her action. 

Summing up, the relational interpretation allows us to introduce the notion of \textit{perspective} in quantum mechanics. In hindsight, this would seem to be a condition sine qua non to make the quantum description applicable to the complex physical situations presented by quantum computation.

The third step is to physically represent the reversibility of the process between the initial and final measurement outcomes. We have argued that this requires evenly sharing the selection of the outcome of the initial measurement (and the consequent selection of that of the final measurement) between the initial and final measurements. To this end, we have assumed that the initial and final measurements reduce to partial measurements that evenly and without redundancy select whatever was selected by the initial measurement in a quantum superposition of all the possible ways of doing this.

The representation of the quantum algorithm with respect to Bob and any external observer, who are not shielded from the outcome of any measurement as usual, remains unaltered under the time-symmetrization in question. That with respect to Alice, who is shielded from the outcome of the initial measurement (the problem setting), changes. Before time-symmetrization, in the input state of the quantum algorithm, Alice is completely ignorant of the outcome of the initial measurement. After it, she knows a part of that outcome corresponding to half of the outcome of the final measurement (the solution) she will read in the future. The complexity of the problem to be solved by her is correspondingly reduced. 

We made the basic assumption that the time-symmetrized representation of the quantum algorithm to Alice, which is the one physically complete, tells us the number of function evaluations required to solve the problem in an optimal quantum way. We have seen that it must be the number required to solve the reduced problem classically. In other words, an optimal quantum algorithm would require the number of function evaluations required by a classical algorithm endowed with the advanced knowledge of half solution. We have called this ``the advanced knowledge rule''.

Going to the discussion, the first thing to say is that the present step is evidently inspired by time-symmetric quantum mechanics $\left [19 \div 26\right ]$. In particular, the time-symmetrization adopted has been inspired by the work of Dolev and Elitzur $\left [23\right ]$ on the non sequential behavior of the wave function highlighted by partial measurement. In both cases there is non locality in time at play: Alice's advanced knowledge of part of a future measurement outcome in our case, the non sequential behavior of the wave function in the other.

A second thing is that the present form of time-symmetrization would seem to bring to completion the notion of the logical $\left [27\right ]$ and physical $\left [28 \div 30\right ]$ reversibility of computation that so much contributed to the development of quantum computation.

Now we discuss some possible objections to the present step. 

 One could reject to start with the assumption that the number of function evaluations required to solve the problem should be dictated by the time-symmetrized representation of the quantum algorithm to Alice. We have just two points in defense of our assumption: (i) When looking for a fundamental explanation of the speedup, one should reason on the complete physical representation of the quantum algorithm; according to our argument this implies relativizing and time-symmetrizing it. (ii) The explanation obtained quantitatively fits all the quantum algorithms examined; in particular it unifies for the first time the quadratic and exponential speedups. Of course we have entered uncharted waters, but this should be acceptable after so many years without a fundamental explanation of the speedup.

A second objection could be that there has never been the need for time-symmetrizing the representation of a reversible quantum process. The answer is that the time-symmetrization in question leaves the ordinary representation unaltered; the need of it comes out only in the case that the observer is shielded from the outcome of the initial measurement. This is a new situation brought about by the extension of the representation of the quantum algorithm to the process of setting the problem.

Another question is the reason for choosing this particular kind of time-symmetrization. We answer by recapping our line of thinking. 

In the case that there is a unitary transformation between the initial and final measurement outcomes, sharing the selection of the initial measurement outcome between the initial and final measurements is a mathematically legitimate operation. Evenly sharing it avoids a time-asymmetry that would be unjustified in a reversible process. Having to take the quantum superposition of all the possible ways of sharing all the possible problem settings at first sight might seem complicated and a drawback. However, this difficulty vanishes when one realizes that such a superposition is the unitary part of the quantum algorithm to Alice back again. Thus, the time-symmetrization in question turns out to be just an interpretation of the internal structure of a quantum superposition, seen as a superposition of time-symmetrized elements along which there is information going back in time from the outcome of the final measurement to that of the initial measurement (we should keep in mind that there is a unitary transformation between the two outcomes). This information is completely redundant with that coming from the initial measurement to an observer not shielded from the outcome of it, it is not when the observer is shielded. In the latter case it tells her a part of the initial measurement outcome corresponding to half of the final measurement outcome she will read in the future.

The net result of completing the physical representation of quantum algorithms is the above said advanced knowledge rule. Besides quantitatively fitting all the optimal quantum algorithms examined, this rule would allow to ascertain the number of function evaluations required to solve any oracle problem in an optimal quantum way.

\subsection{
Synthetic versus analytical approach
}
A general criticism that can be moved to the present explanation of the speedup is that it is not derived mathematically, but from physical interpretations which can always be controversial. As a matter of fact, the most important results about the speedup achieved so far have been derived in an entirely mathematical way by quantum computational complexity theory. The aim of this theory is to find lower and/or upper bounds to the quantum computational complexity of classes of problems and position them with respect to the known classical classes. There is an important body of literature on it, we provide $\left [31\right ]$ as an example.

We would answer the above criticism by an analogy with Euclidean geometry. Geometrical properties can be derived either synthetically, in an axiomatic way from Euclid's postulates, or analytically, in an entirely mathematical way. The present explanation of the speedup, derived from fundamental physical considerations, would be synthetic. The derivations of quantum computational complexity theory would be analytical. Under this analogy, the synthetic and the analytical approaches should not be mutually exclusive. They could be alternative ways of deriving the same properties. 

In principle, there should be an analytical counterpart to the advanced knowledge rule. The number of function evaluations required to solve the problem in an optimal quantum way should be a property of maximum of the generic input-output unitary evolution. Whether the present advanced knowledge rule could be a lead for the search of its analytical counterpart should be for further study.

We should also mention the fact that a synthetic explanation of the speedup, independent of the mathematical complexity of quantum algorithms, has been clearly advocated by Lov Grover in 2001 $\left [32\right ]$. Quoting his words:

\textit{It has been proved that the quantum
search algorithm cannot be improved at all, i.e. any quantum mechanical algorithm will need at least }$\ensuremath{\operatorname*{O}} \left (\sqrt{N}\right )$\textit{ steps to carry out an exhaustive search of }$N$\textit{ items. Why is it not possible to search in fewer than }$\ensuremath{\operatorname*{O}} \left (\sqrt{N}\right )$\textit{ steps? The arguments used to prove this are very subtle and mathematical. What
is lacking is a simple and convincing two line argument that shows why one would expect this to be the case.}

In the context of the present work, a two line argument could be:

\textit{In quantum problem-solving, all is as if the problem solver knew in advance half of the solution she will read in the future.}

Comparing the number of function evaluations foreseen by the advanced knowledge rule with the upper and lower bounds of computational complexity theory should be the object of further investigation. For the time being, we can say that the bounds identified until today are far from exhausting all the kinds of oracle problems. The advanced knowledge rule, instead, given any oracle problem, provides the number of function evaluations required to solve it in an optimal quantum way. If right, let us say, it would be a breakthrough. It would solve a long standing open problem of quantum computer science, the so called quantum query complexity problem (in fact, given any oracle problem, providing tight upper and lower bounds to the number of function evaluations -- oracle queries -- required to solve it in an optimal quantum way).

By the way, the present theory of the speedup, relying on the physical representation of the notion of black box, is limited to oracle quantum computing. This form of computation covers the majority of the quantum algorithms known today but not all of them. A recent approach for developing quantum algorithms is that of the quantum walk algorithms $\left [33\right ]$. Comparing the present approach with this formulation of quantum algorithms might be an interesting object of further investigation.

Back to oracle computing, the notion that an optimal quantum algorithm requires the number of function evaluation of a classical algorithm endowed with the advanced knowledge of half solution should also be a powerful tool for identifying problems liable of interesting speedups.

Another element of comparison between the synthetic and the analytic approach is that the latter does not seem to say anything about the physics involved in the speedup. The synthetic approach instead necessarily provides a bridge between quantum computation and the foundations of quantum mechanics.

\subsection{
Other approaches
}
Another historical approach to the speedup is the investigation of its relation to quantum entanglement. As first noted in $\left [34\right ]$, if the state of the quantum computer register remained a product of the states of its individual qubits throughout the quantum algorithm, this latter could be efficiently simulated (in polynomial time) by a classical algorithm. This means that entanglement  (i. e. non factorizability) must be a distinctive feature of the exponential speedup $\left [35\right ]$ However, the converse, that entanglement implies exponential speedup, may not be true. Until now, no unifying relation between entanglement and speedup could be found. Reference $\left [36\right ]$, after reviewing the role played by entanglement in a variety of quantum algorithms, reaches the following negative conclusion:
\textit{we should give up looking for a single reason behind
the quantum speedup. Most likely, the answer will intimately
be connected with the exact nature of the problem
and, as seen above, will vary from problem to problem.
Though possibly intellectually displeasing, this answer is
the only possible consistent one at present.}

Aside from the fact that the present explanation of the speedup seems to involve a form of temporal entanglement (corresponding to the problem-solution correlation) and a consequent temporal non locality (Alice's advanced knowledge of half of the solution she will read in the future), we are unable to pinpoint any precise relation between it and quantum entanglement -- see also Section 4.4.

We should however note that the explanation in question does not vary from problem to problem. Advanced knowledge of half solution provides a unified way of seeing the speedup. In Grover's problem, knowing in advance half solution (half of the $n$ bits that specify the number of the drawer with the ball) yields obviously a quadratic speedup, from  $\ensuremath{\operatorname*{O}}\left (2^{n}\right )$ to $\ensuremath{\operatorname*{O}}\left (2^{n/2}\right )$ function evaluations. In the problem of finding the period of a function (Simon's and Shor's problems), knowing in advance a good half table \textit{}(half of the table of a period of the function such that there is never repetition of the same value of the function), obviously allows to identify the period with just one function evaluation (for any value of the argument outside that half table) against an exponential number thereof in the classical case. 

Still about the relation between the present work
and literature, we should add what follows:

\begin{itemize}
\item  The possibility that a future action affects a past measurement outcome has been analyzed in
$\left [26\right ]$. Investigating the
relation between
$\left [26\right ]$
and the present work might be an interesting prospect.

\item  The possible relevance of the problem-solution symmetry of Grover algorithm to
the end of explaining its speedup has been hypothesized by Morikoshi
$\left [37\right ]$. In the present work, problem-solution symmetry has been
seen under the more general category of time-symmetry.

\item
We have seen that relational quantum mechanics offers a capability of representing complex physical situations related to a plurality of interrelated observers. Its application to the EPR situation is an interesting precedent $\left [11\right ]$. The physical representation of the fact that either one of the two space-separated observers needs time to receive the measurement outcome obtained by the other observer allows of an unprecedented clarification of the fundamental aspects involved in this situation.

\item
The present explanation of the speedup sheds light on the intersection between two areas of research,
the foundations of quantum mechanics and quantum information, that is currently receiving increasing
attention -- see
$\left [38\right ]$.
\end{itemize}

\begin{itemize}
\item  The advanced knowledge rule, that an optimal quantum algorithm requires the number of function evaluations required by a classical algorithm endowed with the advanced knowledge of half solution, exactly fits all the major quantum algorithms and is very simple. Independently of the argument that supports it, it could be considered an empirical rule for searching prospects of speedup. The search of such rules is another way of approaching the speedup. For example, reference $\left [39\right ]$ highlights a de facto relation (holding for a number of exponential speedups) between quantum computational complexity and a measure of the complexity of the quantum state.\end{itemize}

\subsection{
Cross fertilization prospects}

We have derived a quantitative explanation of the speedup from fundamental physical considerations. There might be a reciprocity. The quantum computational perspective might shed light on the foundations of quantum mechanics.

In particular, the present explanation of the speedup might say something on the issue of causality. It could push forth the notion of \textit{mutual causality} between the initial and final measurement outcomes, as follows. 

For simplicity, let us refer to Grover algorithm, where the problem setting (the number of the drawer with the ball) coincides with the solution. In the ordinary representation of the quantum process, the number in question is selected by the initial measurement; in this case one says that the initial measurement outcome (the problem setting) causes the final one (the solution). In the time reversed picture, the final measurement outcome causes the initial one. In the case that the initial and final measurements reduce to partial measurements that evenly contribute to that selection, in all possible ways in quantum superposition, it can be said that there is \textit{mutual causality} between the initial and final measurement outcomes.

One could also look for possible commonalities between the present temporal form of non locality (Alice's advanced knowledge) and the non locality related to quantum entanglement$\text{}$. Both seem to be related to a notion of mutual causality between respectively time and space separated measurement outcomes -- see  $\left [40\right ]$ for the latter case.

The present form of time-symmetrization shows that there can be information coming back in time along the elements of a quantum superposition. This might shed light on the very nature of quantum superpositions. This observation, which is due to Finkelstein $\left [41\right ]$, could be the object of further investigation.

\subsection{
Conclusion}
We have highlighted a possible connection between quantum computation and the foundations of quantum mechanics.

At the fundamental end, we are simply dealing with a unitary transformation between an initial and a final measurement outcome. Time-symmetrizing this reversible quantum process leaves its ordinary representation, where no observer is shielded from any measurement outcome, unaltered. It changes that relativized to an observer shielded from the outcome of the initial measurement. Immediately after the initial measurement, this observer knows all the same a part of its outcome corresponding to half of the outcome of the final measurement she will read in the future.

At the other end, we have quantum oracle computing. Here the initial measurement outcome is the setting and the final measurement outcome the solution of the problem. Alice, the problem solver, is shielded from the problem setting which is hidden inside the black box. All the same, before beginning her problem solving action, she knows a part of it corresponding to half of the solution she will read in the future. As a consequence, the number of function evaluations required to solve the oracle problem in an optimal quantum way is that of a classical algorithm endowed with the advanced knowledge of half solution. This quantitatively explains the special efficiency of quantum computation, namely the quantum computational speedup.

The explanation in question fits all the quantum algorithms examined, which comprise the major ones, and would solve the quantum query complexity problem.

From the technical standpoint, future work might be checking the present explanation  on a larger basis of existing quantum algorithms, applying it to identify oracle problems liable of interesting speedups and group them into quantum computational complexity classes, comparing these classes with those already identified by quantum computer science.

From the foundational one, future work might be further studying the fundamental implications of the present explanation of the speedup. 

Being based on unconventional physical considerations, the present explanation of the speedup is liable of controversy. Of course we went into uncharted waters. Our defense is that this should be acceptable in the present situation. Apart from the present unconventional approach, up to now there is neither a quantitative nor a fundamental explanation of the quantum computational speedup.

\section*{
Acknowledgments}
Thanks for useful discussions and comments are due to Eli Cohen, Artur Ekert, Avshalom Elitzur, David Finkelstein, Daniel Shehan, and Ken Wharton.

\section*{
References
}
 $\left [1\right ]$ Grover, L. K.: A fast quantum mechanical algorithm for database
search. Proc. 28th Annual ACM Symposium on the Theory of Computing. ACM press New York 212-219 (1996)

$\left [2\right ]$ Ekert, A. K.: Quantum cryptography based on Bell's theorem. Phys. Rev. Lett., 661 (1991)

$\left [3\right ]$ Castagnoli, G. and Finkelstein, D. R.: Theory of the quantum speedup. Proc.
Roy. Soc. A 1799, 457, 1799-1807 (2001) 

$\left [4\right ]$ Castagnoli, G.: The
quantum correlation between the selection of the problem and that of the solution sheds light on the mechanism of the quantum speed up. Phys. Rev. A 82,
052334-052342 (2010) 

$\left [5\right ]$ Castagnoli, G.: Highlighting the mechanism of the quantum speedup
by time-symmetric and relational quantum mechanics. Found. Phys. Vol. 46, Issue 3. 360--381 (2016) 

$\left [6\right ]$ Castagnoli, G.: On the relation between quantum computational speedup
and retrocausality. Quanta Vol. 5, No 1, 34-52 (2016) 

$\left [7\right ]$ Castagnoli, G.: A retrocausal model of the quantum computational
speedup. Proceedings of the 92nd Annual Meeting of the Pacific Division of the American Association for the Advancement
of Science, Quantum Retrocausation III, Program organizer Daniel Sheehan (2016)

$\left [8\right ]$ Deutsch, D.: Quantum theory, the Church Turing principle and the
universal quantum computer. Proc. Roy. Soc. A 400, 97-117 (1985)

$\left [9\right ]$ Rovelli, C.: Relational
Quantum Mechanics. Int. J. Theor. Phys. 35, 637-658 (1996) 

$\left [10\right ]$ Rovelli, C.: Relational Quantum Mechanics (2011) http://xxx.lanl.gov/pdf/quant-ph/9609002v2

$\left [11\right ]$ Rovelli, C. and Smerlak, M.: Relational EPR. Preprint: arXiv:quant-ph/0604064

$\left [12\right ]$ Bennett, C. H., Bernstein, E., Brassard, G., and Vazirani, U.:
Strengths and Weaknesses of Quantum Computing. SIAM Journal on Computing Vol. 26 Issue 5, 1510-1523 (1997)

$\left [13\right ]$ Long, G. L.: Grover algorithm with zero theoretical failure rate.
Phys. Rev. A 64, 022307-022314 (2001) 

$\left [14\right ]$ Toyama, F. M., van Dijk, W., and Nogami Y.: Quantum search with
certainty based on modified Grover algorithms: optimum choice of parameters. Quantum Information Processing 12, 1897-1914 (2013)

$\left [15\right ]$ Deutsch, D. and Jozsa,
R.: Rapid solution of problems by quantum computation. Proc. Roy. Soc. A 439, 553-558 (1992)

$\left [16\right ]$ Simon, D.: On the power of quantum computation. Proceedings of
the 35th Annual IEEE Symposium on the Foundations of Computer Science 116-123 (1994) 

$\left [17\right ]$ Kaye, P., Laflamme, R., and Mosca, M.: An Introduction To Quantum
Computing. Oxford University Press 146-147 (2007) 

$\left [18\right ]$ Shor, P.: Algorithms for quantum computation: Discrete log and factoring. Proceedings of
the 35th Annual IEEE Symposium on the Foundations of Computer Science 124-131 (1994)

$\left [19\right ]$ Wheeler, J. A. and Feynman, R. P.: Interaction with the Absorber
as the Mechanism of Radiation. Rev. Mod. Phys. 17, 157 (1945)

$\left [20\right ]$ Watanabe, S.: Symmetry of physical laws. Part III. Prediction and retrodiction. Reviews of Modern Physics. 27 (2), 179-186 (1955)

$\left [21\right ]$ Aharonov, Y., Bergman,
P. G., and Lebowitz, J. L.: Time Symmetry in the Quantum Process of Measurement. Phys. Rev. B 134, 1410-1416
(1964)

$\left [22\right ]$
Cramer, J.: The Transactional Interpretation of Quantum Mechanics.
Rev. Mod. Phys. 58, 647 (1986)

$\left [23\right ]$ Dolev, S. and Elitzur,
A. C.: Non-sequential behavior of the wave function. arXiv:quant-ph/0102109 v1 (2001)

$\left [24\right ]$ Aharonov, Y. and Vaidman, L.: The Two-State Vector Formalism: An
Updated Review. Lect. Notes Phys. 734, 399-447 (2008) 

$\left [25\right ]$ Aharonov, Y., Popescu, S., and Tollaksen,
J. A.: Time-symmetric formulation of quantum mechanics. Physics Today, November issue 27-32 (2010)

$\left [26\right ]$ Aharonov, Y., Cohen, E., and Elitzur, A. C.: Can a future choice affect a past measurement outcome? Ann. Phys. 355, 258-268 (2015)

$\left [27\right ]$ Fredkin, E. and Toffoli, T.: Conservative Logic. Int. J. Theor. Phys.
21, 219-253 (1982) 

$\left [28\right ]$ Landauer, R.: Irreversibility and heat generation in the computing
process. IBM Journal of Research and Development, 5 (3), 183--191, 1961

$\left [29\right ]$ Finkelstein, D. R.: Space-time structure in high energy interactions.
In Gudehus T, Kaiser G, Perlmutter A editors, Fundamental Interactions at High Energy. New York: Gordon \& Breach, 1969 pp. 324-338

$\left [30\right ]$ Bennett, C. H.: The Thermodynamics of Computation -- a Review. Int.
J. of Theor, Phys 21, 905-940 (1982)

$\left [31\right ]$
Aaronson, S. and Ambainis, A.: Forrelation: a Problem that Optimally Separates Quantum from Classical Computing. arXiv:1411.5729 (2014)

$\left [32\right ]$
Grover, L. K.: From Sch{\"o}dinger's Equation to the Quantum Search
Algorithm. arXiv:quant-ph/0109116 (2001) 

$\left [33\right ]\text{}$ Venagas-Andraca, S. E.:Quantum walks: a comprehensive review. arXiv:1201.4780 (2012)

$\left [34\right ]$ Jozsa, R.: Entanglement and Quantum Computation. Geometric Issues
in the Foundations of Science, eds.: Huggett, S., Mason, L. K. P., Tod, L. K. P., Tsou, S. T.,
and Woodhouse N. M. J., Oxford University Press (1997)

$\left [35\right ]$ Jozsa, R. and Linden, N.: On the role of entanglement in quantum
computational speed-up. Proc. Roy. Soc. A 1097 (2002) 

$\left [36\right ]$
Vedral, V.: The elusive source of quantum effectiveness. \ref{}\ref{Issue 8}.

$\left [37\right ]$ Morikoshi, F.: Problem-Solution Symmetry in Grover's Quantum Search
Algorithm. Int. J. Theor. Phys. 50, 1858-1867 (2011)

$\left [38\right ]$ Workshop on Quantum Foundations and Quantum Information -- Theory and Experiment. . Yakir Aharonov, Avshalom Elitur, and Eliahu Cohen organizers. https://indico.cern.ch/event/559774/page/8707-workshop-on-quantum-foundations-and-quantum-information

$\left [39\right ]$ (2015)

$\left [40\right ]$ Price, H. and Wharton K.: Disentangling the Quantum World. arXiv:1508.01140v2 (2015)

$\left [41\right ]$ Finkelstein, D. R. Private communication.

\end{document}